\documentclass[journal,twoside,web]{ieeecolor}
\usepackage{generic}
\usepackage{cite}
\usepackage{amsmath,amssymb,amsfonts}
\usepackage{algorithmic}
\usepackage{graphicx}
\usepackage{textcomp}
\def\BibTeX{{\rm B\kern-.05em{\sc i\kern-.025em b}\kern-.08em
    T\kern-.1667em\lower.7ex\hbox{E}\kern-.125emX}}
\usepackage[acronym,automake]{glossaries}
\makeglossaries
\newacronym{amp}{AMP}{amplifier}
\newacronym{ap}{AP}{action potential}
\newacronym{dft}{DFT}{discrete Fourier transform}
\newacronym{DPS}{DPS}{discrete, prolate-spheroidal sequences}
\newacronym{eeg}{EEG}{electroencephalogram}
\newacronym{fft}{FFT}{{fast Fourier transform}}
\newacronym{fir}{FIR}{finite impulse response}
\newacronym{hfd}{HFD}{high-frequency demodulate}
\newacronym{hfh}{HFH}{HFD harmonic}
\newacronym{hfl}{HFL}{high-frequency LFP}
\newacronym{hfn}{HFN}{HFD noise}
\newacronym{hlm}{HLM}{high-frequency LFP / MUA}
\newacronym{hmc}{HMC}{harmonic}
\newacronym{iid}{IID}{independent and identically distributed}
\newacronym{inp}{INP}{input}
\newacronym{int}{INT}{integrated}
\newacronym{ios}{IOS}{input-output system}
\newacronym{lfl}{LFL}{low-frequency LFP}
\newacronym{lfp}{LFP}{local field potential}
\newacronym{llh}{LLH}{LFL harmonic}
\newacronym{lln}{LLN}{LFL noise}
\newacronym{mea}{MEA}{microelectrode array}
\newacronym{mua}{MUA}{multiunit activity}
\newacronym{nrm}{NRM}{normal mode}
\newacronym{nse}{NSE}{noise}
\newacronym{oup}{OUP}{Ornstein-Uhlenbeck process}
\newacronym{phs}{PHS}{physical output}
\newacronym{ptf}{PTF}{power transfer function}
\newacronym{rem}{REM}{rapid eye movement}
\newacronym{res}{RES}{residual}
\newacronym{sgn}{SGN}{signal}
\newacronym{snr}{SNR}{signal-to-noise ratio}

\begin{document}
\title{Temporal-lobe Epilepsy: Harmonic and Anharmonic Periodicity in Microeletrode Voltage}
\author{Fran\c{c}ois A. Marshall, \IEEEmembership{Member, IEEE}
\thanks{Submitted \today. This work was supported by National Institute of Neurological Disorders and Stroke R01NS110669. }
\thanks{F. A. Marshall completed this work as Postdoctoral Associate (2020-2022) and Visiting Scholar (2022) with Mathematics and Statistics Boston University, 111 Cummington Mall \#140C, Boston, MA 02215, United States (e-mail: francois.marshall@queensu.ca).}
}

\maketitle

© 20xx IEEE. Personal use of this material is permitted. Permission from IEEE must be obtained for all other uses, in any current or future media, including reprinting/republishing this material for advertising or promotional purposes, creating new collective works, for resale or redistribution to servers or lists, or reuse of any copyrighted component of this work in other works.

\begin{abstract}
Temporal-lobe epilepsy in humans is often associated with widespread, synchronized neuron firing that co-occurs with traveling waves in local field potential. These traveling waves generate stochastic oscillations in a time series of microelectrode voltage, and previous work has deemed it informative for traveling-wave analysis to study the mean periodicity. This manuscript reveals that: a) mean voltage (i.e., traveling-wave periodicity) adequately explains the observed voltage periodicity only for a select few time intervals during seizure; and b) mean voltage has a 7\,Hz cosine-series representation indicative of a nonlinear system response given alpha-rhythm input. The a) result implies that residual noise should be modelled explicitly, while b) motivates a departure from the conventional plane-wave modeling regime in source-localization efforts. The 7\,Hz fundamental frequency is unsurprising given the relative transparency of the brain to 14\,Hz alpha rhythms in neurophysiological diseases (14\,Hz being a subharmonic frequency of the 7\,Hz signal).
\end{abstract}


\section{Introduction}
\label{sec:introduction}

\IEEEPARstart{T}{he} electrophysiological manifestation of temporal-lobe epilepsy is high-synchronization spatiotemporal organization of neocortical multiunit firing. When multiunit activity is outstanding, synaptic potential also exhibits stereotyped behaviour: producing traveling waves in \acrfull{lfp}. This manuscript considers voltage recordings from a \acrfull{mea} in a deep-lying neocortical layer\,\footnote{Specific details of this setup are discussed in \cite{schlafly2022multiple}.}. In previous literature\,\footnote{e.g., \cite{diamond2021travelling,martinet2017human,schlafly2022multiple}.}, the phase of an ictal-discharge traveling wave is used to localize a spherical-wave source.; that is, the signal is assumed a wavepacket\,\footnote{What \cite{hecht1998optics} describes as a narrowband superposition of normal modes that destructively interfere to make the resultant appear as if the superposition of two equal-amplitude, different-frequency plane waves.}. In \cite{martinet2017human}, the spatiotemporal \acrshort{lfp} model is refined: the traveling-wave periodicity is thought to approximate steady-state dynamics of a Turing-Hopf dynamical system with stochastic forcing. To this end, the wavepacket represents time-evolutionary mean \acrshort{mea} voltage whose normal modes have high \acrfull{snr}.

This manuscript introduces a cosine-series model for mean \acrshort{mea} voltage. In assuming a nonlinear system function, the new model departs from the wavepacket modeling convention because: a MacLaurin polynomial whose argument is a neural rhythm\,\footnote{Here, a neural rhythm is thought to be a nonrandom, narrowband $L^2$-signal that has a cosine-series expansion.} can have large enough order that the normal modes of the system-response function are highly dispersed across the nonnegative half of the principal domain. An advantage of the new model is that it admits nonlinearity and stochastic periodicity that are realistic qualities of the underlying cortical dynamics. To this end, the identified normal modes are less likely to be spurious signal elements than in the wavepacket model. This manuscript discusses the results of a novel multitaper spectral analysis of MEA-voltage data, accounting for the time evolution of voltage periodicity during the course of a human seizure.

All code and results used for the analysis of this manuscript can be found at the code repository of \cite{marshall2023harmonicgitHub}. There at that repository can be found a useful \texttt{README.docx} file, which itself makes reference to that important instructions document which is referenced here as \cite{marshall2023codealgorithmsgitHub}. The remainder of this section presents the model. Then Section\,\ref{sec:results_section} presents a longitudinal analysis for one human-subject / \acrshort{mea}-electrode element of the complex survey. The analysis reveals: how the periodicity of mean voltage specifies a partition of the voltage record during seizure into four epochs (Section\,\ref{sec:figure0_section}); and convincing evidence for a 7\,Hz fundamental frequency of the cosine series (Section\,\ref{sec:period_estimation_mean_signal_ACVF_appendix}). Finally: the discussion proposes a nominal biophysical mechanism to explain the source of periodicity for two of the epochs, while motivating further work to understand mechanisms responsible for the other two epochs (Section\,\ref{sec:discussion_section}).

For ${\Delta t>0}$ a sampling period and ${x:\mathbb{R}\to\mathbb{R}}$, the time series, ${\mathbf{x}\in\mathbb{R}^N}$, of MEA-voltage is given by
\begin{equation}
\mathbf{x}^{\mathrm{T}} = [\, x(t_n) \,]_{n=0}^{N-1},
\label{eq:time_series_vector_equation}
\end{equation}
where, for ${n\in\mathbb{Z}}$: the $n$'th time, $t_n$ is given by
\begin{equation}
t_n = n\cdot\Delta t.
\label{eq:tn_equation}
\end{equation}
Model $\mathbf{x}$ by the generative random vector, $\mathbf{X}$, where\,\footnote{For \ref{eq:vector_X_continuous_X_relationship_equation}, the notation of $T_N^{-1}$ in the $X$ argument is a reminder that $t_n$ enters as ${t_n\cdot T_N^{-1}}$ in any timelimiting function - the custom for Fourier analysis, as discussed in\,\cite{brillinger1981time}.
}
\begin{equation}
\mathbf{X}[n]=X\left(\,\frac{t_n}{T_N}\,\right),
\label{eq:vector_X_continuous_X_relationship_equation}
\end{equation}
with the terms given as follows.
\begin{enumerate}
\item $X$ is a real-valued $L^p$-process on $\mathbb{R}$\,\footnote{Refer to \cite{krishnan1984nonlinear} for details about $L^p$ stochastic processes.}, where ${p\in\mathbb{N}_{>1}}$.\,\footnote{Justification for the choice of $p$ is provided in \cite{marshall2020advances}.}
\item $T_N$ is the record length, where ${T_N = t_{N-1}}$.
\end{enumerate}
The MEA-voltage has the signal-plus-noise decomposition,
\begin{equation}
X = x^{(SGN)} + X^{(NSE)},
\label{eq:LFP_HFD_decomposition_equation}
\end{equation}
where $x^{(SGN)}$ and $X^{(NSE)}$ respectively denote the mean \acrfull{sgn} and \acrfull{nse}. Denote by $J_{SGN}$ the order of the cosine series.

In this manuscript, the data considered comprise a 30\,kHz voltage series from a single MEA electrode and have been drawn from a complex clinical survey of neocortical seizure activity. Details of the survey can be found in \cite{schlafly2022multiple}, while the data are available at \cite{schlafly_kramer_human_seizure_2022}. The electrode having unit identification index is considered from the Subject\,C5 study (refer to Table\,1 in \cite{schlafly2022multiple} for details).

\section{Results}
\label{sec:results_section}

\subsection{The Time-evolutionary Periodicity of Microelectrode Voltage}
\label{sec:figure0_section}

\begin{figure}[!t]
    \centerline{\includegraphics[width=\columnwidth,trim=0 290 300 0,clip]{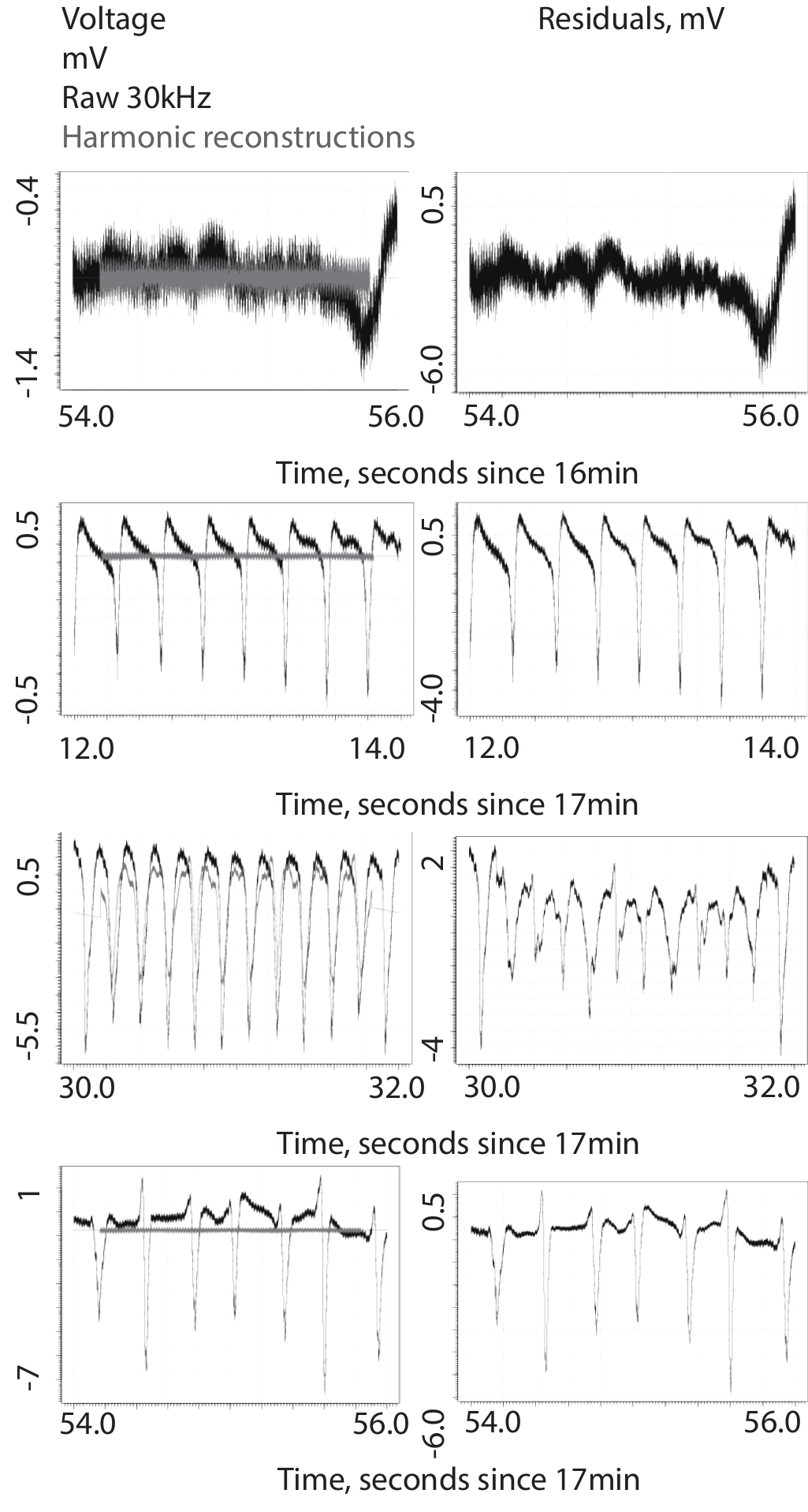}}
\caption{A cosine-series signal accurately represents mean MEA-voltage in some epochs but not others. Left: a 30\,kHz section trace (black) overlaid by a cosine-series reconstruction (grey). Right: residuals of the cosine-series reconstruction.}
\label{fig:Figure0}
\end{figure}

For a survey of time sections, preliminary MEA-voltage multitaper power-spectral and harmonic analyses reveal for $x^{(SGN)}$-reconstructions how a sparse cosine-series signal can explain the $X$ mean signal. Fig.\,\ref{fig:Figure0} presents the following.
\begin{enumerate}
\item \underline{First column}:
\begin{itemize}
\item $\mathbf{x}$ (black, ${|\Delta t|^{-1} = 3\times10^4}$, ${N=6\times10^4}$).
\item An $x^{(SGN)}$-reconstruction (grey)\,\footnote{The reconstructions have been obtained using a modified version of the 50\% section-overlap method of \cite{marshall2019multitaper} (in particular, the same choice of multitaper time-bandwidth and zero-padding parameters). The modifications regard: the estimation of mode frequencies; and linear extrapolation of this nominal reconstruction at the record ends. The modifications and general reconstruction algorithm are all detailed in Algorithm\,3 of \cite{marshall2023codealgorithmsgitHub}. The reconstruction is denoted by $\check{x}^{(SGN)}$ in Section\,IV\,A of \cite{marshall2023codealgorithmsgitHub}. For ${N=6\times10^4}$, the size of the grid of the zero-padded fast Fourier transform is ${1.31072\times10^5}$. Refer to Section\,II of \cite{marshall2023codealgorithmsgitHub} for details of these parameter initializations.
}.
\end{itemize}
\item \underline{Second column}: A reconstruction of the expected value of
\begin{equation*}
X^{(NSE)}\,\|\,\{X=x\}.
\end{equation*}
\end{enumerate}

In each plot row of Fig.\,\ref{fig:Figure0}, the 30\,kHz trace ($\mathbf{x}$) exhibits periodicity stereotypical of one of four epochs. Table\,\ref{table_epochs} displays how each plot row indicates the dominance of one or the other of $x^{(SGN)}$ and $X^{(NSE)}$ during the considered section window. The table includes rows entitled ``Alpha periodicity'' because - in the corresponding epochs - the fundamental frequencies of the $x^{(SGN)}$ cosine series are 7.2\,Hz and 21.9\,Hz, respectively; both are approximately mode frequencies of a 7\,Hz cosine series\,\footnote{Details of the frequency calculations are specified in the section forthcoming; for now, remark that 14\,Hz - a multiple of 7\,Hz - is the alpha-band upper bound specified by \cite{shi2019relationship}.
}.
Yet stronger evidence for the 7\,Hz hypothesis is that the third plot row shows approximately 12 full troughs occurring between 30.0 and 31.8\,s - and 12 cycles in 1.8\,s yields 6.7\,Hz. As for the ``Ictal wavefront'' and ``Ictal discharges'' epoch names, these have been chosen based on the terminology of \cite{schevon2019multiscale}. The table also includes columns for the base time and the time interval of that plot row of Fig.\,\ref{fig:Figure0} which the table row details. Mean voltage accurately explains the observed $\mathbf{x}$-periodicity (first and third plot rows, with $x^{(SGN)}$ included in the ``Dominant'' column of Table\,\ref{table_epochs}) whenever the grey $x^{(SGN)}$-reconstruction trace: tracks the black $x$-trace; and has comparable amplitude to that of the $x$-trace fluctuations. In cases where $x^{(SGN)}$ does not so well explain $x$ (i.e., second and fourth plot rows of Fig.\,\ref{fig:Figure0}), Table\,\ref{table_epochs} identifies $X^{(NSE)}$ as the main contributing $X$-component. In other words, it may well be possible to make the assumption that $X^{(NSE)}$ has nonzero mean signal and then add more of those modes to $x^{(SGN)}$ in the revised model; however, this would lead to a cosine series that is less sparse and so having higher estimation variance\,\footnote{Details are provided in \cite{marshall2020advances} about constructing a sparse cosine-series model for the autocorrelation function of $X^{(NSE)}$ - typically, a more efficient signal-processing method than simply adding more normal modes to $x^{(SGN)}$ whenever $x$ contains sharp-edged waveforms.}.

\begin{table}[h]
\caption{The Epochs of an MEA-voltage Record}
\label{table}
\setlength{\tabcolsep}{3pt}
\begin{tabular}{|p{100pt}|p{20pt}|p{40pt}|p{40pt}|}
\hline
&&&\\
\textbf{Epoch} & \textbf{Base} & \textbf{Interval} & \textbf{Dominant} \\
& \textbf{min} & \textbf{s} &\\\hline
&&&\\
Alpha periodicity, early seizure & 16 & [54.0,56.0] & $x^{(SGN)}$ \\\hline
&&&\\
Ictal wavefront & 17 & [12.0,14.0] & $X^{(NSE)}$ \\\hline
&&&\\
Alpha periodicity, late seizure & 17 & [30.0,32.0] & $x^{(SGN)}$ \\\hline
&&&\\
Ictal discharges & 17 & [54.0,56.0] & $X^{(NSE)}$ \\\hline
\end{tabular}
\label{table_epochs}
\end{table}

Fig.\,\ref{fig:Figure0} shows how the epochs of MEA-voltage during a human seizure can be identified simply using the cosine-series signal element of mean voltage. For the first row, $\mathbf{x}$-points produce: a dense, black, central region; and a sequence of local minima and maxima more resolved in time. Absence of these minima and maxima in the corresponding residuals plot reveals these points to belong to $x^{(SGN)}$. For the third plot row, the smooth curvature of $x$ in a time interval between sharp-trough events explains the good fit of $x^{(SGN)}$ (the grey and black curves line up). By contrast, remark the sharp $x$ peaking that occurs in the second and fourth plot rows - it is said that $X$ is anharmonic (i.e., its variability dominated by that of $X^{(NSE)}$) during the epochs of ictal wavefront and ictal discharges (second and fourth plot rows).

\begin{figure}[!t]
    \centerline{\includegraphics[width=\columnwidth,trim=300 290 0 0,clip]{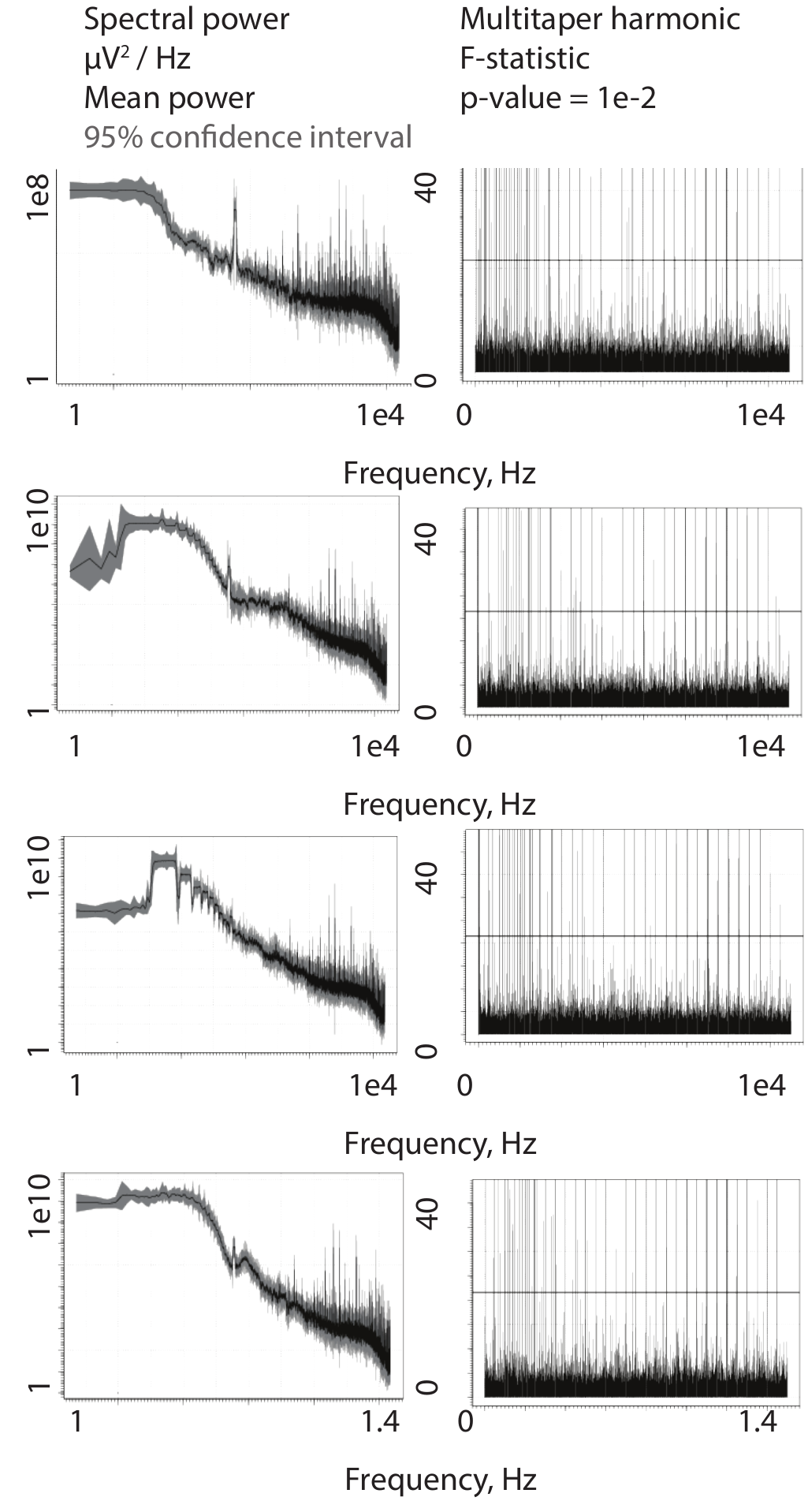}}
\caption{The amplitudes and frequencies of mean-voltage normal modes are revealing about generative oscillations.
Left: spectral power - jackknifed estimates of mean (black) and 95\% confidence bounds (bounded grey region) of the multitaper spectral-power estimator - point-by-point across the non-negative part of the principal domain. Right: spectra of multitaper harmonic F-statistic values (99.998\% the Bonferroni level, i.e. ${10^{-2}}$ p-value).}
\label{fig:Figure1}
\end{figure}

Plotting the frequency-domain signals that characterize the finite-dimensional distribution of $X$ provides useful diagnostics to explain the functional behaviours observed in the plots of Fig.\,\ref{fig:Figure0}. To this end, consider the diagnostic plots of Fig.\,\ref{fig:Figure1}: whose four plot rows correspond to those of Fig.\,\ref{fig:Figure0}, and whose contents are as follows\,\footnote{The power-spectrum reconstructions have been obtained using the multitaper estimation scheme discussed in \cite{marshall2020advances}.
}.
\begin{enumerate}
\item \underline{First column}:
\begin{itemize}
\item Spectral power - multitaper jackknife mean estimates (black, \cite{thomson2007jackknifing}).
\item Multitaper 95\% confidence intervals (grey, \cite{thomson2007jackknifing}).
\end{itemize}
\item \underline{Second column}:
\begin{itemize}
\item Multitaper F-statistic spectrum\,\cite{thomson1990quadratic}.
\item Bonferoni ${\left(\,1-\frac{1}{N}\,\right)}$ level - horizontal line\,\footnote{Refer to \cite{marshall2020advances} for details about the Bonferoni level. Algorithm\,1 includes a data-adaptive step for test-threshold assignment that ensures at least a user-specified number (3 the default) of identified $x^{(SGN)}$ normal modes, but since the Bonferoni threshold satisfies the 3-detection criterion the threshold adaptation was unncessary.
}.
\end{itemize}
\end{enumerate}

\noindent\\
During the early-seizure alpha and ictal-wavefront epochs, the power spectrum (first and second plots, respectively) displays three notable features, as follows.
\begin{enumerate}
\item A 60\,Hz spectral peak (alternating current\,\footnote{Refer to Section\,6.2.2 in \cite{kramer2016case}.}). This is the tallest peak in both plots, occurring midway along the frequency axis.
\item Conspicuous high-frequency spectral peaks right of the 60\,Hz peak.
\item A log-linear decay indicative of red noise\,\footnote{Refer to \cite{fougere1985accuracy,solo1992intrinsic} for a description of red noise.}
\end{enumerate}

\noindent
As for the late-seizure alpha and ictal-discharge epochs, remark how the 60\,Hz spectral peak disappears, but that the log-linear trend and high-frequency peaks remain similar to what they are for the other two epochs.

In the second column of Fig.\,\ref{fig:Figure1}, the F-statistic spectra reveal a plethora of level-${99.99}$\% peaks. The reason that the peak frequencies extend all the way out to the 30\,kHz Nyquist rate is because no one of the considered time sections has pure-sinusoid periodicity; for example, even given the accurate fit in the third plot row of Fig.\,\ref{fig:Figure0}, there remains significant residual periodicity (refer to the residuals plot right of the 30\,kHz-trace / fit overlay). That F-statistic peak height does not decay prior to the Nyquist rate is evidence that alpha periodicity enters the system as an alpha rhythm in the argument of the nonlinear system function.

\subsection{Period Estimation for the Mean Signal}
\label{sec:period_estimation_mean_signal_ACVF_appendix}

\begin{figure}[!t]
    \centerline{\includegraphics[width=\columnwidth,trim=0 0 0 0,clip]{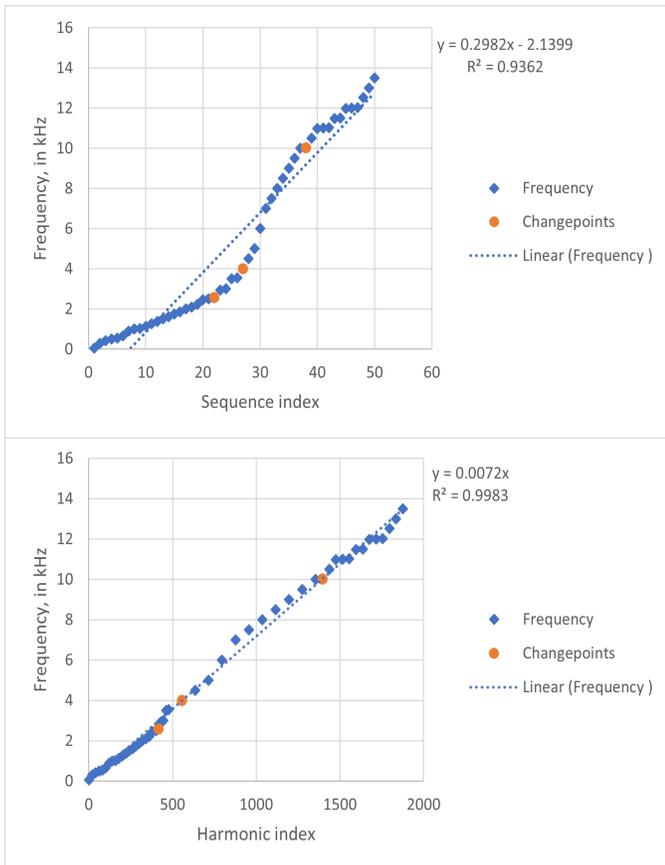}}
\caption{Between them, mean-voltage mode frequencies in the early-seizure alpha epoch reveal a 7.2\,Hz fundamental frequency.
Top: Identified mode frequency vs. sequence index, given the training section that corresponds to the first row in Table\,\ref{table_epochs}. Bottom: Same as the top, but now with scale expansions to sequence-index step size (resulting in harmonic indices, changepoints correspond to scale changes). Display box: fitted trendline model and the $R^2$-value.}
\label{fig:FigureA2}
\end{figure}

The multitaper F-statistic spectra in Fig.\,\ref{fig:Figure1} revealed normal modes in $x^{(SGN)}$. For each of the early- and late-seizure alpha epochs, this section reveals further that the collection of identified normal modes corresponds to those of just a single cosine series - as opposed to the alternative, which is a sum of indepedent cosine series: each having unique fundamental frequency. In addition, the inferred fundamental frequency of $x^{(SGN)}$ in either epoch suggests the presence of an alpha-rhythm source. Table\,\ref{table_A1} displays both results and performance diagnostics of the estimation algorithm for fundamental frequency\,\footnote{For a detailed overview, refer to Algorithm\,4 in \cite{marshall2023codealgorithmsgitHub}.
}. For fundamental-frequency inference the early-seizure alpha epoch (Table\,\ref{table_A1}, first row), both the table contents and estimation method are now described making use of Fig.\,\ref{fig:FigureA2}, which itself includes the following.
\begin{enumerate}
\item \textbf{Top} Scatter plot (solid, blue diamonds) of mode frequency versus sequence index (the sequence indices being
\begin{equation*}
\{1,2,\cdots,\check{J}_{SGN}\},
\end{equation*}
with $\check{J}_{SGN}$ the number of identified mode frequencies and an integer estimate of $J_{SGN}$ - ``No. ID'' in Table\,\ref{table_A1}). Solid, orange discs and table entries are discussed in the second list item below. Linear correlation occurs across a number of scatter-point clusters, as illustrated with the use of the plotted linear trendline (remark from the plot insert that ${R^2=0.94}$).
\item \textbf{Bottom} The same plot as the top one, except now with the following two changes so that all scatter points approximately correspond to just a single cosine series.
\begin{enumerate}
\item Dynamic scale expansion of the spacing between pairs of sequence indices (corresponding to harmonic indices - that is: integral multiples of a single fundamental frequency).
\item A zero-intercept constraint in the linear regression of identified frequencies on the harmonic indices.
\end{enumerate}
Up to a sequence index of 21, the scatter points of the top plot (solid, blue diamonds) exhibit high linear correlation. So too, do the scatter points having sequence indices between 22 and 26 (the 22nd scatter point is a changepoint for the slope; in both top and bottom plots a solid, orange disc is overlaid on the solid, blue diamond). To align both of these two scatter-point clusters, different trial spacings are introduced between sequence indices to produce trial harmonic indices. As seen in the bottom plot on either side of the orange disc: a 20-step spacing suffices (the 1(20) entry under the Table\,\ref{table_A1} ``Changepoint (Steps)'' column). Now, the first two clusters combined exhibit strong linear correlation; however, neither this cluster nor the cluster of remaining scatter points are collinear with the resulting linear trendline. Thus: sequence indices of the second cluster in the bottom plot are converted to harmonic indices by raising the sequence-index separation size from 1 to 15 (the 22(15) Table\,\ref{table_A1} entry). This action makes more of the scatter points collinear in total - and so brings the scatter points nearer the trendline on average. Next, the top plot is used to identify a third high-correlation cluster (sequence indices 27 through 37; solid, orange disc overlaid on the 37'th scatter point of both top and bottom plots) inside the larger cluster of remaining scatter points. Then, this third cluster is aligned with the first two of the bottom plot (the 27(80) Table\,\ref{table_A1} entry). The process is continued until all scatter points align both with each other and with the resulting trendline (remark from the plot insert that ${R^2=0.998}$).
\end{enumerate}

\begin{table}[h]
\caption{Fundamental-frequency Calculations for the Mean-voltage Cosine Series}
\label{table}
\setlength{\tabcolsep}{3pt}
\begin{tabular}{|p{15pt}|p{37pt}|p{35pt}|p{25pt}|p{45pt}|p{25pt}|}
\hline
&&&&&\\
\textbf{Base} & \textbf{Interval} & \textbf{Frequency} & \textbf{No.\,ID} & \textbf{Changepoint} & {$\pmb{R^2}$} \\
\textbf{min} & \textbf{s} & \textbf{Hz} &  & \textbf{(Steps)} & \\\hline
&&&&&\\
16 & [54.0,55.8] & 7.2 & 65 & 1(20), 22\,(15), 27\,(80), 38\,(40) & 0.9983\\\hline
&&&&&\\
17 & ${[30.0,31.8]}$ & 21.9 & 60 & 1(1), 7(20), 10\,(5), 24(15), 33(20) & 0.9953\\\hline
\end{tabular}
\label{table_A1}
\end{table}

\begin{figure}[!t]
     \centerline{\includegraphics[width=\columnwidth,trim=0 0 0 0,clip]{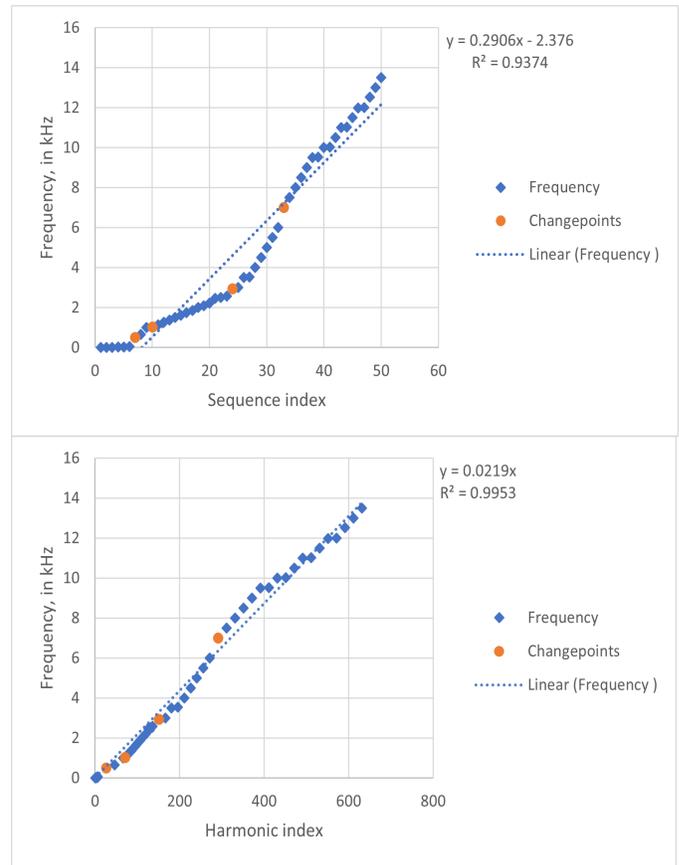}}
\caption{Between them, mean-voltage mode frequencies in the late-seizure alpha epoch reveal a 21.9\,Hz fundamental frequency.
The same as Fig.\,\ref{fig:FigureA2}, except now with the training section corresponding to the third row in Table\,\ref{table_epochs}.}
\label{fig:FigureA3}
\end{figure}

\noindent
Fig.\,\ref{fig:FigureA3} displays the same contents as Fig.\,\ref{fig:FigureA2}, except now for the late-seizure alpha epoch. The second row of Table\,\ref{table_A1} contains the relevant estimates and performance diagnostics. Given the high $R^2$-value, it would appear that a single cosine series is adequate to explain the periodicity in mean voltage.

\section{Discussion}
\label{sec:discussion_section}

In Table\,\ref{table_A1}, 21.9\,Hz is approximately three times the 7.2\,Hz fundamental frequency; therefore, both of the associated cosine series have a 7\,Hz fundamental frequency. The alpha band - ${(9,14]}$\,Hz\,\cite{shi2019relationship} - contains the basebands of several oscillations in temporal-lobe\,\footnote{Here, an oscillation is distinguished from a rhythm because the former is just an $L^2$-demodulate of the generative, random neural data signal - there is no constraint about it being nonrandom or narrowband.
}, extracellular, electric potential whose spatial range is appreciable. Alpha oscillations are attributed to the following connectivity dynamics during human seizure, and long range of the oscillations indicates that the temporal lobe during epileptic seizure is relatively transparent to 14\,Hz oscillations.
\begin{enumerate}
\item During medial refractory seizures, \acrshort{mea} recordings reveal how the ${(4,14]}$\,Hz oscillation accurately predicts \acrfull{ap} timing\,\cite{zanos2012relationships}\,\footnote{For a complex survey of MEA-recordings including three cell strata, the number of cases where \acrshort{lfp} successfully predicts spike timings falls from 26 down to 8 when shifting from the ${(4,14]}$\,Hz band to the ${(14,30]}$\,Hz band.}
\item Both during the wake and natural-sleep brain states, the ${[10,20]}$\,Hz-oscillation is excited in the mesial temporal lobe \cite{uchida2001cortical}. In addition, \cite{uchida2001cortical} lists the following references that sighted a limbic-system 14\,Hz normal mode during \acrfull{rem} sleep: \cite{malow1999hippocampal,montplaisir1981sleep,nakabayashi2001absence}.
\item Recent literature specifies a spindle as an $x^{(SGN)}$-mode whose line power is approximately the average spectral mass in the ${[10,15]}$\,Hz sigma frequency band during States\,2,3 non\acrshort{rem} sleep\,\footnote{Refer to Section\,3.2.7 in \cite{spencer2021biomarker}. Section\,3.3.1 in that reference specifically considers: \cite{beelke2000relationship,nobili1999relationship,nobili2001temporal,
nobili2001distribution, tucker2009impact}}. In \cite{uchida2001cortical}, 14\,Hz is deemed characteristic of sleep spindles\,\footnote{The ``spindle'' term was coined in \cite{loomis1935potential}, and the phenomenon first identified in \cite{berger1933Elektroencephalogram} - see \cite{de2003sleep} for an overview.
}: citing Fig.\,6 of \cite{brazier1968studies}, in which occurs an outstanding 14\,Hz spectral peak in simultaneous electrode recordings of both the hippocampal gyrus and interior\,\footnote{Remarkable evidence for this 14\,Hz power is also encountered in Fig.\,5 of the more recent work, \cite{de2003sleep}: in the frontal and parietal cortices, coincident 14\,Hz peaks occur during non\acrshort{rem} sleep.}. Whereas \cite{brazier1968studies} itself presents no evidence of 14\,Hz spindles in \acrfull{eeg} recordings\,\footnote{In more recent works, Section\,3.3.1 in \cite{spencer2021biomarker} states that the following works have indeed observed in \acrshort{eeg} recordings large-scale propagation via an association between spindle voltage and sigma power: \cite{de2003sleep,wamsley2012reduced}.
}, it does provide other evidence for long-range connectivity.
\end{enumerate}

\noindent
In \cite{summer2022sleep}, it is suggested that the spike-and-wave discharges in epilepsy during non\acrshort{rem} sleep might be generated in ways similar to spindles\,\footnote{In particular, refer to the spindle / spike-and-wave discharge comparison in Fig.\,1 of \cite{leresche2012sleep}.}: in the same spirit for proposing biophysical models under a common-source hypothesis, it is here proposed that the mechanism responsible for the inferred 7\,Hz oscillation arises in part due to the presence of an alpha rhythm. A candidate mechanism is the action of a nonlinear \acrshort{mea} system function on a 7\,Hz wavepacket; \cite{feynman2010lectures} shows how the result is a 7\,Hz cosine series\,\footnote{The order, $J_{SGN}$, depends on how many derivatives the nonlinear system function has at the origin - a consequence of MacLaurin series expansions. The greater the number of derivatives, the smoother the system function.
}.

\section{Conclusion}
This manuscript has revealed that the sparse cosine-series modeling of mean MEA-voltage fails to explain the outstanding periodicity during epochs of ictal wavefronts and of ictal discharges. As such, rigorous dynamical modeling should include nonlinear and/or stochastic noise specification. It appears that the epochs of early- and late-seizure alpha periodicity both share a common 7\,Hz source (suggesting the presence of a source alpha rhythm), but that a 7\,Hz rhythm appears only as a driver for the MEA system function. This hypothesis precludes spatiotemporal localization of the source because more about the system function must be known before asserting that the plane-wave dynamics of a single source are generative of mean MEA-voltage.

\section*{Acknowledgment}

F.A. Marshall thanks Prof. Mark Kramer, Ms. Emily Schlafly and Dr. Ani Wodeyar\,\footnote{Computational Neurodata and Modeling Lab, Mathematics and Statistics, Boston University.} for their contributions in discussions regarding the underlying theory and biophysics of cortical traveling waves received by \acrshort{mea}'s.

\printglossary[type=\acronymtype]

\bibliographystyle{plain}
\bibliography{Bibliography}

\begin{IEEEbiography}[{\includegraphics[width=1in,height=1.25in,clip,keepaspectratio]{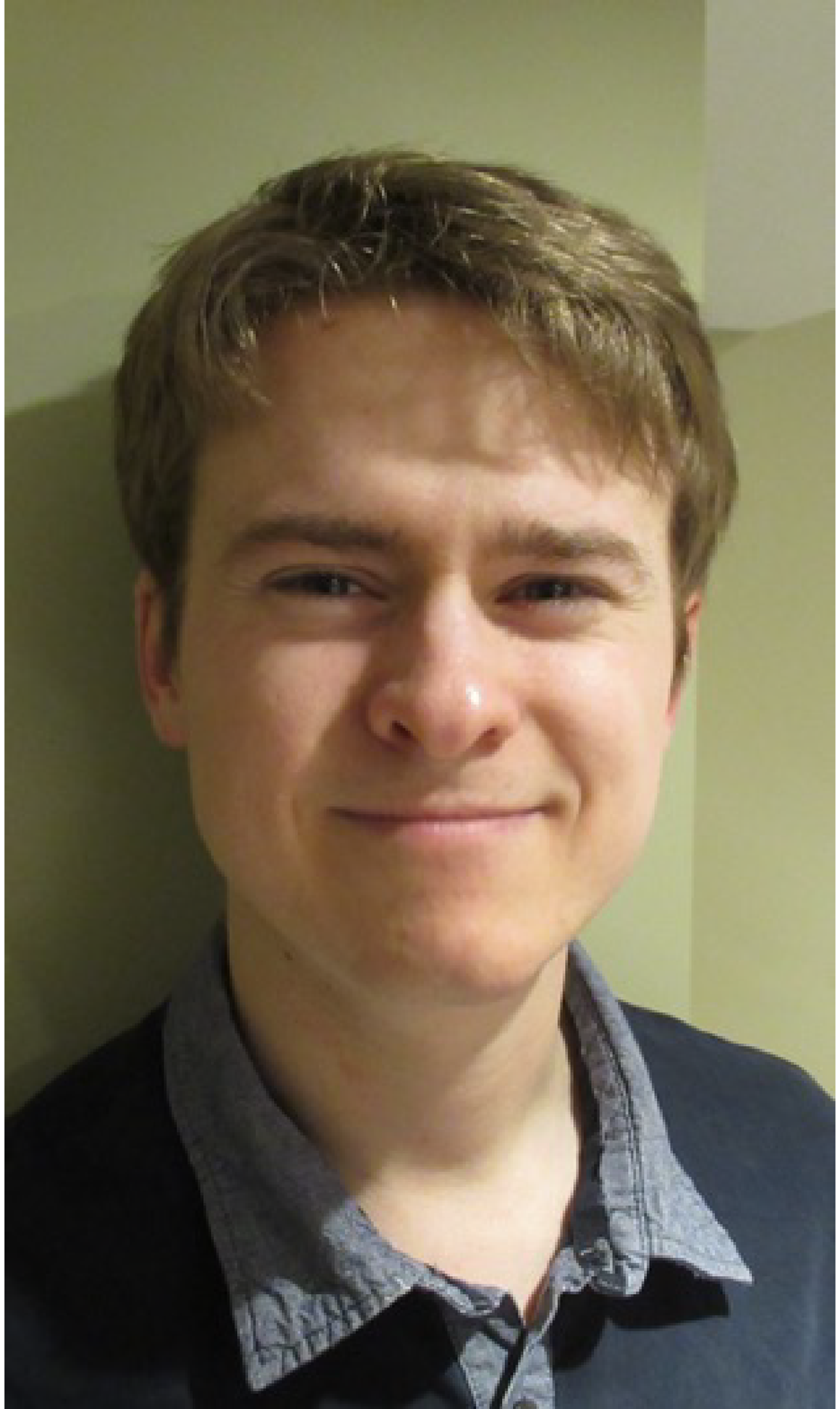}}]{Fran\c{c}ois A. Marshall} was born in London, England in 1990. He received: B.S. (Theoretical Physics, 2012) and M.S. (Medical Imaging Physics, 2014) degrees from Carleton University, Ontario, Canada; and a Ph.D. degree (Applied Mathematics, 2020) from Queen's University, Ontario, Canada.

During the M.Sc., he was Research Assistant with Natural Resources Canada (Nuclear Emergency Response Team). In Winter\,2019, he was Teaching Fellow for ``STAT\,464/864 Time Series Analysis'' at Queen's University. From 2020 to 2022, he was a Postdoctoral Associate with Boston University Mathematics and Statistics. From October to December, 2022: he was a Visiting Scholar at Boston University (Neuromodeling and Data Lab, Mathematics and Statistics). He is first author on: \cite{marshall2018multitaper,marshall2019acharacterization}; and second author on \cite{sinclair2015reconstruction}. His research interests include: time-series and spectral analyses; stochastic processes and Monte Carlo; probability theory and analysis; and neuroscience.

Dr.\ Marshall is Secretary of Statistical Society of Canada (SSC) Student and Recent Graduates Committee and Secretary of SSC Probability Section. He is a member of: American Epilepsy Society; American Geophysical Union; Royal Statistical Society; and SSC (Statistical Society of Ottawa). In 2019, he was a recipient of the SSC Probability Section Student Research Presentation Award.

\end{IEEEbiography}

\end{document}